\documentclass[conference]{IEEEtran}
\usepackage{fancyhdr}
\IEEEoverridecommandlockouts
\usepackage{makecell}
\usepackage{multirow}
\usepackage{multicol}
\usepackage{subfigure}
\usepackage{listings}
\usepackage{amsmath}
\usepackage{pifont}

\usepackage{cite}
\usepackage{amsmath,amssymb,amsfonts}
\usepackage[ruled,vlined]{algorithm2e}  

\usepackage{graphicx}
\usepackage{textcomp}
\usepackage{booktabs}
\usepackage{xcolor}
\def\BibTeX{{\rm B\kern-.05em{\sc i\kern-.025em b}\kern-.08em
    T\kern-.1667em\lower.7ex\hbox{E}\kern-.125emX}}
\begin{document}

\title{ 

MCFuser: High-Performance and Rapid Fusion of Memory-Bound Compute-Intensive Operators
}

\author{

\IEEEauthorblockN{Zheng Zhang}
\IEEEauthorblockA{\textit{School of Computer Science} \\
\textit{Wuhan University} \\
zzhang3031@whu.edu.cn}
\and

\IEEEauthorblockN{Donglin Yang}
\IEEEauthorblockA{\textit{NVIDIA} \\
dongliny@nvidia.com}
\and

\IEEEauthorblockN{Xiaobo Zhou}
\IEEEauthorblockA{
\textit{University of Macau} \\
waynexzhou@um.edu.mo}
\and

\IEEEauthorblockN{Dazhao Cheng\IEEEauthorrefmark{1}} \thanks{\IEEEauthorrefmark{1}Corresponding author.}
\IEEEauthorblockA{\textit{School of Computer Science} \\ \textit{Wuhan University} \\
dcheng@whu.edu.cn}
\and

}

\maketitle

\thispagestyle{fancy}
\lhead{}
\rhead{}
\chead{}
\lfoot{\footnotesize{
SC24, November 17-22, 2024, Atlanta, Georgia, USA
\newline 979-8-3503-5291-7/24/\$31.00 \copyright 2024 IEEE
}
}
\rfoot{}
\cfoot{}
\renewcommand{\headrulewidth}{0pt}
\renewcommand{\footrulewidth}{0pt}

\begin{abstract}
Operator fusion, a key technique to improve data locality and alleviate GPU memory bandwidth pressure, often fails to extend to the fusion of multiple compute-intensive operators due to saturated computation throughput. However, the dynamicity of tensor dimension sizes could potentially lead to these operators becoming memory-bound,
necessitating the generation of fused kernels — a task hindered by limited search spaces for fusion strategies, redundant memory access, and prolonged tuning time, leading to sub-optimal performance and inefficient deployment.

We introduce MCFuser, a pioneering framework designed to overcome these obstacles by generating high-performance fused kernels for what we define as memory-bound compute-intensive (MBCI) operator chains. Leveraging high-level tiling expressions to delineate a comprehensive search space, coupled with Directed Acyclic Graph (DAG) analysis to eliminate redundant memory accesses, MCFuser streamlines kernel optimization. 
By implementing guidelines to prune the search space and incorporating an analytical performance model with a heuristic search, MCFuser not only significantly accelerates the tuning process but also demonstrates superior performance. Benchmarked against leading compilers like Ansor on NVIDIA A100 and RTX3080 GPUs, MCFuser achieves up to a 5.9x speedup in kernel performance and outpaces other baselines while reducing tuning time by over 70-fold, showcasing its agility.

\end{abstract}


\section{Introduction}

As deep learning models grow in size and complexity, they necessitate a myriad of operators, which in turn significantly burdens GPU memory bandwidth with frequent I/O operators~\cite{antman, whale,mpmoe,reddgnn}. Operator fusion emerges as a pivotal strategy to mitigate this challenge~\cite{chimera,dnnfusion,astitch}, aiming to diminish kernel launch overhead and improve data locality by curtailing the reliance on frequent global memory access. It has been widely embraced across various libraries~\cite{cublas,cudnn,cutlass,tensorrt,mpipemoe} and tensor program compilers~\cite{ansor,tvm,tiramisu,astra,taso,scalable-kernel-fusion,akg,tensor-comprehension,xla} to enhance runtime performance by consolidating operators, thereby optimizing computational efficiency. Frameworks such as Cutlass~\cite{cutlass}, TensorRT~\cite{tensorrt}, and high-level compilers like TVM~\cite{tvm,autotvm} and Ansor~\cite{ansor}, have deployed operator fusion to address computational and memory bandwidth constraints.

Deep learning operators typically fall into one of two categories: compute-intensive, like GEMM, or memory-intensive, such as LayerNorm and Softmax. The latter often saturates GPU memory bandwidth due to their frequent memory accesses, resulting in the under-utilization of computing resources. To address this, several studies have explored operator fusion, either among memory-intensive operators or between memory-intensive operators and a single compute-intensive operator, aiming to optimize resource use. 
Specifically, AStitch~\cite{astitch} focuses solely on fusing memory-intensive operators, while DNNFusion~\cite{dnnfusion} fuses a compute-intensive operator with its adjacent memory-intensive operators. Despite these efforts, the fusion of compute-intensive operator chains, which could further exploit computing resources, remains largely unaddressed. This oversight reveals a significant opportunity for enhancing computational performance beyond the current scope of most deep-learning compilers.

The distinction between computation and memory bandwidth bottlenecks in deep learning operators is not solely dependent on the operator type but also on input tensor characteristics. This is demonstrated by how dynamic changes in tensor dimensions can shift an operator's nature. For example, in a GEMM operator $C_{M,N}=A_{M,K}\times B_{K,N}$, altering the dimension $K$ from 1,024 to 1 on modern accelerators can dramatically change the computation to memory access ratio from 227 to 2, transitioning the operator from compute-bound to memory-bound. In scenarios where reduction dimensions are minimized, traditionally compute-intensive operators can unexpectedly encounter memory bandwidth limitations, transforming them into memory-bound scenarios. This observation leads us to identify a unique category of operators, termed memory-bound compute-intensive (MBCI) operators, which exhibit this dual nature under certain conditions.

MBCI operators, exemplified by self-attention modules in transformative architectures like the Transformer~\cite{transformer}, underscore a critical optimization frontier in deep learning. For instance, in the BERT-base model processing a 1,024-sequence input, self-attention constitutes only 14\% of the computational workload (FLOPs) but dominates 51\% of the execution time, revealing substantial optimization potential. Despite the promising avenue for performance enhancement through kernel fusion, represented typically by hierarchical loop structures and computation-memory access patterns, several obstacles impede the efficient fusion of MBCI operator chains. Firstly, the search space for fusion strategies often remains incomplete; for example, Chimera~\cite{chimera} limits itself to nested loop computations, neglecting alternate execution sequences within the same loop scope. Secondly, the direct coupling of memory accesses with computation loops can lead to redundant data movements, particularly noticeable in cases where loop extents are minimized, as seen in Chimera's implementation. Lastly, the fusion process is frequently bogged down by prolonged auto-tuning phases and unwieldy search spaces. 
Frameworks like Ansor require hours for operator optimization through ML-guided exploration (e.g., XGBoost~\cite{xgboost}), relying on extensive runtime statistics for model training and significantly increasing the tuning overhead.

To overcome the outlined challenges in fusing MBCI operator chains, we introduce MCFuser, a framework designed for high performance and efficiency. MCFuser starts with constructing a comprehensive search space using high-level tiling expressions, enabling the complete exploration of potential fused operator decompositions. By leveraging DAG analysis, we strategically decouple memory access statements from loops, significantly reducing redundant memory transfers between GPU global and shared memory. To streamline the search process and mitigate time expenditures, MCFuser employs heuristic guidelines for pruning non-viable segments of the search space. A pivotal innovation of our framework is the development of an analytical performance model based on an exhaustive analysis of computational and memory transactions. This model dramatically cuts down the need for extensive program measurement and model training. Incorporating this performance model into a heuristic search mechanism, MCFuser automatically identifies the optimal configuration for fused tensor programs, marking a significant leap forward in the efficient optimization of MBCI operators.

In summary, our contributions are as follows:
\begin{itemize}
    \item We identify opportunities for enhancing data locality in tensors by analyzing compute-intensive operator bottlenecks across varying input tensor dimensions, which lead to the fusion of MBCI operators.
    \item We construct a comprehensive search space for MBCI operator fusion through detailed tiling expressions and reducing redundant memory access via DAG analysis.
    \item We develop efficient pruning guidelines and integrate an analytical performance model with a heuristic search method, which collectively streamlines the tuning process, enabling automatic and efficient identification of the optimal candidate in the search space. 
\end{itemize}

We encapsulate our innovations in MCFuser and rigorously evaluate its performance across diverse workloads on NVIDIA GPUs. MCFuser's tensor programs significantly surpass existing methods, achieving up to a 5.9× speedup over Ansor~\cite{ansor} and consistently outperforming competitors like BOLT~\cite{bolt} and PyTorch~\cite{pytorch}. Moreover, it dramatically reduces tuning times by more than 70 times, underscoring MCFuser's efficiency and effectiveness in optimizing tensor operators.

\section{Background and Motivation}
\label{sec:background}

\subsection{Operator Fusion Characteristics}
\label{sec:motivation-challenges}

Operator fusion, acclaimed for reducing kernel launch and memory access overheads, categorizes tensor operators into compute-intensive and memory-intensive types. Compute-intensive operators, such as GEMM, demand high computational resources, whereas memory-intensive operators, like ReLU and softmax, are limited by memory bandwidth~\cite{astitch,chimera}. This classification underlies optimization strategies, notably epilogue fusion, which merges a compute-intensive operator with the following memory-intensive ones, illustrated by the coupling of MatMul and ReLU~\cite{cutlass,bytetransformer,raptort}. These practices aim to balance computational demands with memory efficiency.

\begin{figure}
\centering
\scalebox{0.8}{
\centerline{\includegraphics[trim=1 1 0 5,clip]{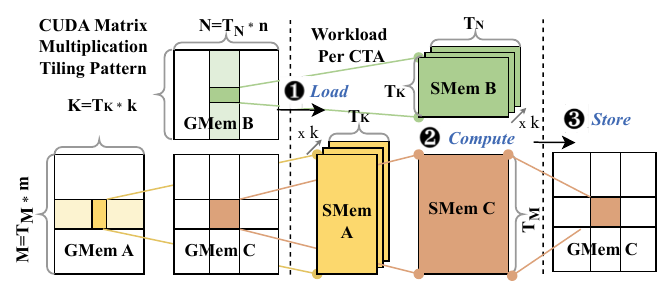}}
}
\setlength{\abovecaptionskip}{0cm}
\caption{ 
Illustration of the Matmul operator implemented on the CUDA platform, indicating \textit{GMem} for global memory and \textit{SMem} for shared memory.
}\label{fig:gemm}
\setlength{\belowcaptionskip}{0cm}
\end{figure}

However, while compute-intensive operators are generally characterized by high computational intensity, they can become memory-bound as tensor dimensions change. Consider the MatMul ($C=A\times B+C$) operator as an example. Illustrated in Fig.~\ref{fig:gemm}, each thread block computes a tile of C, initially loading tiles from global memory to shared memory~\ding{182}, then accumulating partial results of tile $C$ over $k$ iterations~\ding{183}, and finally writing the result back to global memory~\ding{184}.

\begin{figure}
\centering
\scalebox{0.75}{
\centerline{\includegraphics[trim=5 5 10 0,clip]{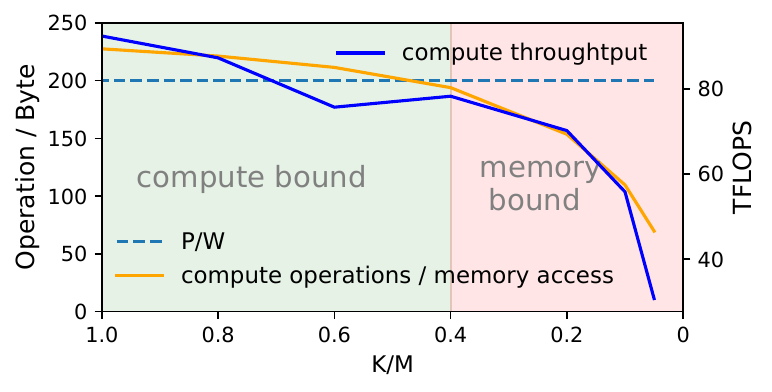}}
}
\setlength{\abovecaptionskip}{-0.1cm}
\caption{
The performance of a MatMul operator across different K/M ratios while maintaining constant computational complexity ($M\times N\times K=1024^3$), where $M$ = $N$). The theoretical ratio of compute operations to memory accesses for a tile size of 256 is represented on the left y-axis, while the actual computation throughput measured on the A100 GPU is shown on the right y-axis.
}\label{fig:pre}
\end{figure}

The theoretical ratio of compute operations to memory accesses, denoted as $\phi = (2T_MT_NK) / (2T_MT_N + T_MK + T_NK)$, serves as a critical indicator of performance bottlenecks. As illustrated in Fig.\ref{fig:pre}, a decrease in the $K/M$ ratio leads to a higher proportion of memory accesses relative to compute operations, making the tensor program increasingly memory bandwidth-bound and diminishing computational throughput. This dynamic reveals that the compute-intensive operator, MatMul, transitions to a memory-bound state when $\phi < \mathcal{P}/\mathcal{W}$, with $\mathcal{P}$ being the peak computation throughput and $\mathcal{W}$ the memory bandwidth, thus classifying it as an MBCI operator. Similar to memory-intensive operators, fusing MBCI operators improves data locality and mitigates memory bandwidth demands. However, traditional fusion strategies that rigidly classify compute-intensive operators as non-fusible boundaries fail to capitalize on the benefits of fusing MBCI operators, as highlighted by prior studies\cite{astitch, dnnfusion, fusionstitching}.

\begin{table*}[tb]
\centering
\setlength{\abovecaptionskip}{0.1cm}
\caption{The Comparison among existing representative works and MCFuser.} \label{tab:related}
\scalebox{0.93}{
\begin{tabular}{l|l|l|l|l|l|l}
\hline
& \multicolumn{3}{c|}{\bf Search Space} & {\bf Memory Optimize} &\multicolumn{2}{c}{\bf Exploration} \\
\hline
\bf Name & \bf \makecell[l]{Support MBCI } & \bf\makecell[l]{Auto.}&\bf \makecell[l]{Search Space} & - & \makecell[l]{\bf Optimize Objective / Guidance}  & \bf\makecell[l]{Tuning time} \\
\hline
AStitch~\cite{astitch}  & No   &  Yes & Stitch Schemas    & Fusion   &  Rule-based & Short  \\
DNNFusion~\cite{dnnfusion} & No&  Yes & Pattern-based   &   Fusion & Mathematical Analyze &  Short \\ 
BOLT~\cite{bolt}        & Partial  &  Yes &  Template-based & Fusion & Measured Performance   & Mid\\
FlashAttention~\cite{flashattention} & Partial& No&Handcrafted     & Fusion & - & -\\
Ansor~\cite{ansor}      & Yes  & Yes &  Loop Transformation & + Loop Opt. & ML Cost Model(e.g. XGBoost~\cite{xgboost})  &  Long   \\
Chimera~\cite{chimera}  & Yes &Yes  & Nested Block Execution Order & + Loop Opt. & Minimize Data Movement & Short \\
MCFuser~(ours)     & Yes  &Yes &  Exhaustive Tiling-based & + Rid of Redundancy & Analytical Performance Model  & Short \\ 
\hline
\end{tabular}
}
\end{table*}

\textbf{A Typical Case:} Transformer models~\cite{transformer}, pivotal in natural language processing and image classification, grapple with efficiency bottlenecks, especially in their self-attention modules, which exhibit quadratic time and memory complexity relative to sequence length~\cite{flashattention}. For example, in the Bert-Large model by HuggingFace~\cite{huggingface}, despite self-attention modules contributing only 11\%, 14\%, and 19\% of the FLOPs for sequence lengths of 512, 1024, and 2048, respectively, they disproportionately dominate execution time, consuming 39\%, 51\%, and 61\% of it. This stark contrast highlights the self-attention module's inability to fully leverage computing resources, constrained significantly by memory bandwidth, and underscores the need for optimization strategies that address these specific inefficiencies.

\subsection{Challenges and Limitations of Previous Works}~\label{sec:challenges}
In this section, we identify three primary challenges encountered in the development of efficient kernels for fusing MBCI operator chains. Furthermore, we discuss how existing literature falls short in addressing these challenges, with a comprehensive summary presented in Table~\ref{tab:related}.

\paragraph{Incomplete Search Space} 
A fundamental challenge in fusing MBCI operator chains arises from an incomplete search space. For instance, AStitch~\cite{astitch} focuses on fusing traditional memory-intensive operators like element-wise, broadcast, and reduction operators, including LayerNorm and Softmax, overlooking the fusion potential for MBCI operators. Similarly, DNNFusion~\cite{dnnfusion}  merges compute-intensive operators with adjacent ones but refrains from directly fusing multiple compute-intensive operators, citing complexity. This limitation is echoed in several studies~\cite{bytetransformer,fusionstitching,tiramisu}, which stereotype compute-intensive operators as non-fusible boundaries, missing opportunities for broader operator fusion. While efforts like Ansor~\cite{ansor}, BOLT~\cite{bolt}, Chimera~\cite{chimera}, and FlashAttention~\cite{flashattention, flashattention2} have ventured into the fusion of MBCI operators, they each confront their own search space constraints. Ansor's loop structure scheduling expands the search space but at the cost of efficiency~\cite{hidet}. BOLT facilitates the fusion of dual GEMM operators through template matching yet struggles with more complex modules like self-attention, which are not included in its fusion patterns. FlashAttention optimizes self-attention but requires extensive expert knowledge and lacks versatility for different workloads. Chimera attempts a comprehensive fusion strategy for compute-intensive operators yet is limited by a search space confined to nested loop permutations. The collective limitations of these frameworks highlight a critical gap in existing methodologies, underscoring the necessity for a more adaptive and encompassing search strategy to unlock the full potential of MBCI operator fusion. 

\paragraph{Redundant Memory Access} Operator fusion aims to reduce memory accesses by storing intermediate tensors in shared memory. However, the fusion of multiple compute-intensive operators often leads to redundant memory access, as the coupling of computation blocks and loops may not always optimize memory usage. For example, loops specific to producer operators that also include consumer operators can cause unnecessary repeated accesses to the same memory region. Frameworks like Chimera attempt to optimize memory access by moving statements to the outermost relevant loops, a strategy mirrored in Ansor's {\tt GetLastReduceIteratorInOutermostReduceTile} function. Despite these efforts, both frameworks miss opportunities to minimize access overhead in scenarios where loop extents are trivially one, highlighting a gap in fully leveraging memory access optimization.

\paragraph{Exploration and Tuning Cost} 
The search for the optimal schedule incurs significant time and resource costs across many compilers, such as TVM~\cite{tvm}, Halide~\cite{halide}, and FlexTensor~\cite{flextensor}. For instance, Ansor's extensive search space, navigated via an ML model like XGBoost, demands running statistics from online programs for model training, thereby increasing tuning overhead. While BOLT alleviates some of this burden through its template-based design, it still requires considerable template instantiation and program measurement, adding to the tuning effort. On the other hand, Chimera attempts to lower these costs using an analytical model focused on data transfer optimization. However, by neglecting the computational redundancy, it often arrives at sub-optimal scheduling decisions.

\section{Search Space Generation and Optimization}\label{sec:design-space}
In response to these identified challenges, MCFusor is designed with a comprehensive search space derived from tiling-based expressions. Through the utilization of DAG flow analysis, we efficiently minimize redundant memory access and filter out invalid search candidates, significantly streamlining the exploration process for enhanced efficiency.

\begin{figure}[tb]
\centering
\scalebox{0.475}{
\centerline{\includegraphics[trim=1 1 1 10,clip]{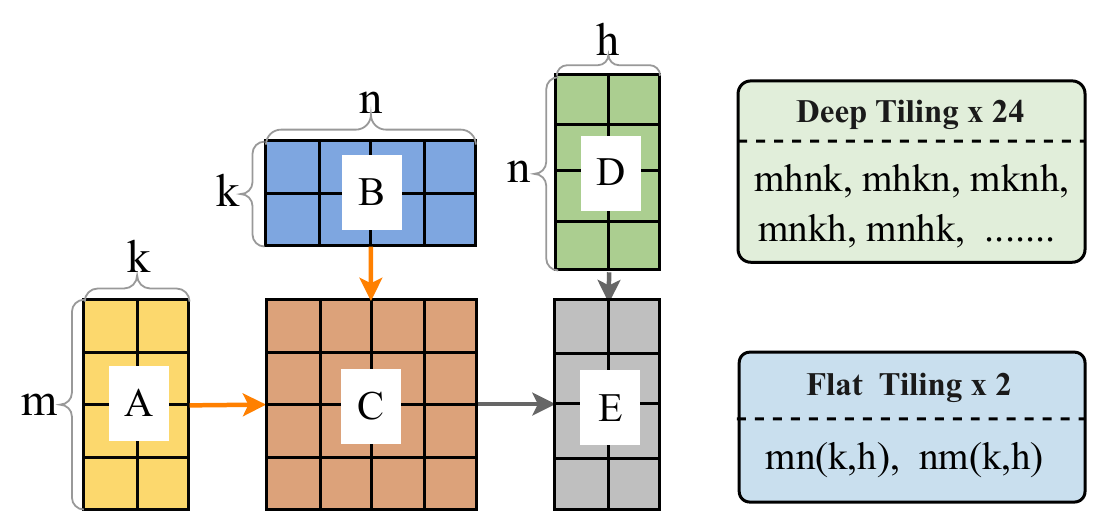}}
}
\setlength{\abovecaptionskip}{0cm}
\caption{ 
The GEMM chain consists of two GEMM operators, with $m, k, n, h$ representing the number of tiles in each dimension.
}\label{fig:gemm-chain}
\end{figure}

\subsection{Search Space Generation}\label{sec:tiling-expression}

In deep learning models, compute-intensive operators can be broken down into computation blocks surrounded by cross-tile loops, where each block iteratively accesses tiles of input tensors~\cite{amos,atomic-dataflow,roller}. These loops are represented by a vector $\overset{\rightarrow}{l}=(l_1,l2,...l_J)$, where $J$ denotes the number of independent loops and $l_j$ signifies the $j$-th loop with $l_j$ representing its extent. Any candidate in the search space can be delineated by the structure of loops and the values of $\overset{\rightarrow}{l}$. 

Initially, we utilize a straightforward tiling expression to delineate the structure of loops. The relationship between loops can be categorized into two types: Nested, where $l_jl_{i}$ signifies that the loop $l_{i}$ is executed within the scope of $l_{j}$.
Sequential, where  $(l_j, l_i)$ indicates that the loops $l_{i}$ and $l_{j}$ are executed sequentially within the same scope.
For clarity, we exemplify with the GEMM chain scenario that comprises two compute-intensive operators. However, our analysis method naturally extends to scenarios with more compute-intensive operators. Additionally, we apply standard fusion optimizations for memory-intensive operators in line with previous work~\cite{ansor,astitch}, although such optimizations are beyond the scope of this paper.

As depicted in Fig.~\ref{fig:gemm-chain}, the GEMM chain ($C=A\times B, E=C\times D$) is composed of four cross-tile loops: $m, n, k, h$, where $m$ and $n$ are shared loops for computation blocks $C$ and $E$. Based on the loop relations outlined above, we categorize all tiling expressions into two categories:

\begin{itemize}
    \item \textbf{Deep Tiling.} In this category, any two loops exhibit a nested relation, and the maximum depth of loops equals the number of loops. If there are  $x$ separate loops, then the number of permutations of loops is $x!$, such as $mnhk, mnkh, mkhn$, etc. For example, in the GEMM chain, this results in $4!=24$ permutations in total.
    \item \textbf{Flat Tiling.} In this category, there exists loops with sequential relation, limiting the maximum depth of loops to be smaller than the number of loops. In the GEMM chain example, two flat tiling expressions exist: $mn(k,h)$ and $nm(k,h)$.
\end{itemize}

For the tile sizes $\overset{\rightarrow}{l}$, given that tensor cores require a minimum tile size of $16\times16\times16$, we consider all $T_x$ that are multiples of 16 and smaller than the dimension size as viable options. Thus, the total number of candidates in our search space is determined by multiplying the number of possible tiling expressions with the viable tile size options.

Informed by our comprehensive analysis of tiling expressions and tile sizes, we are able to fully enumerate the candidates within the search space for fused kernels. Unlike Chimera, which focuses on deep tiling expressions to the exclusion of flat tiling, our approach ensures no potential configuration is overlooked. This distinction is crucial, as evidenced by the sub-optimal results of Chimera's strategy detailed in the experiments of Section~\ref{sec:exp:graph}.

\begin{figure}[tb]

\centering
\subfigure[Memory node $S_E$ moved from within loop $n$ to loop $h$.] {
 \label{fig:mem-opt-code1}     
\includegraphics[width=0.428\columnwidth,trim=0 0 0 0,clip]{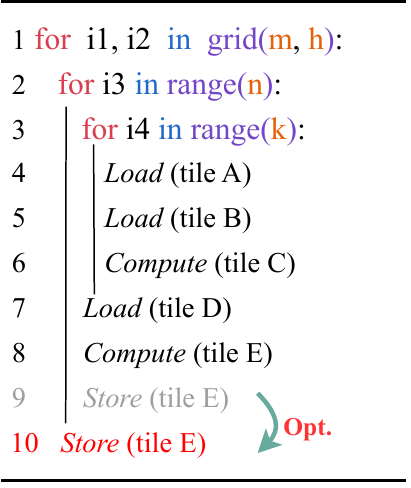}  
}     
\hspace{0.05\columnwidth}
\subfigure[Memory node $L_A$ shifted from within loop $k$ to loop $m~(k=1)$.] { 
\label{fig:mem-opt-code2}     
\includegraphics[width=0.435\columnwidth,trim=0 0 0 0,clip]{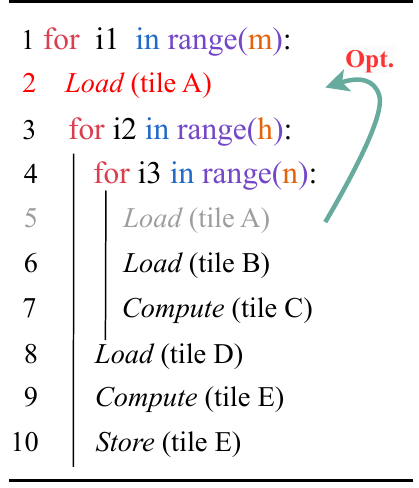}     
}  
\caption{
The pseudo-code that illustrates the optimized tiling expression $mhnk$ for a fused GEMM chain, highlighting memory node optimization.
}\label{fig:tile-exprs}
\end{figure}

\subsection{Memory Access Optimization} 
\label{sec:mem-plan}

We enhance the basic tiling expression, addressing cross-tile loops, by adding details on computation and memory access. It introduces three primitive operations: $Load$ ($L$), $Compute$ ($C$), and $Store$ ($S$), denoted as suffixes attached to each operation, indicating the tensor it operates on. The position of $Compute$ primitive statements is determined by the rightmost related loops. By default, $Load$ and $Store$ primitive statements are associated with the corresponding $Compute$ statement. For instance, considering the GEMM chain depicted in Fig.~\ref{fig:gemm-chain}, the related loops for computation blocks $C$ and $E$ are $m$, $k$, $n$ and $m$, $n$, $h$, respectively. The basic expression of $mhnk$ can be expanded to $mhn(k(L_A, L_B, C_C), L_D, C_E, S_E)$, as illustrated in lines 1-9 of Fig.~\ref{fig:mem-opt-code1}.

 The volume of data movement for each tensor is calculated as the product of tile size and the trip counts of surrounding loops. However, if the loop variables surrounding a certain operation are not utilized for accessing tensor tiles, it results in redundant operations. To mitigate this, memory access statements can be relocated to the rightmost related loop. For example, as depicted in Fig.~\ref{fig:mem-opt-code1}, line 9 shows $S_E$ within the $m$, $h$, $k$ loops, but since $k$ is not used to access tile $E$, moving the $Store(\text{Tile E})$ to the $h$ loop scope, as in line 10, cuts memory accesses by a factor of $k$. 
 This optimization, represented by the tiling expression $mh(n(k(L_A, L_B, C_C), L_D, C_E), S_E)$, reduces memory traffic. 
 Although this method has been adopted by frameworks like Ansor and Chimera, it is often possible to further reduce redundancy, especially when the optimal tile size matches the dimension size, making some loop variables superfluous. In such scenarios, memory access efficiency is improved by adjusting the statements to outer loops, thus decreasing unnecessary trip counts.

\begin{figure}
\centering
\subfigure[The DAG of $mhnk$.] {
 \label{fig:mem-dag1}     
\includegraphics[width=0.8\columnwidth,trim=0 0 0 0,clip]{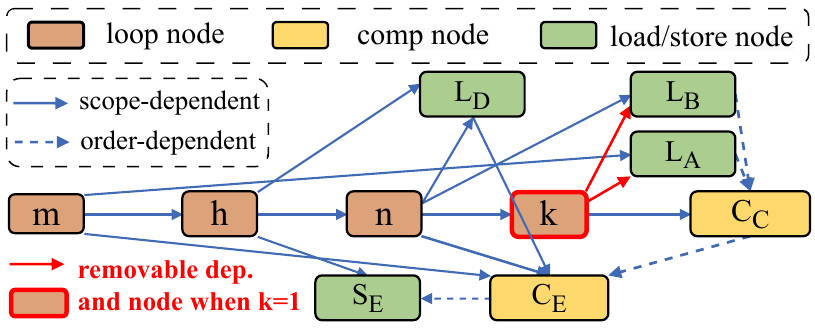}  
}     
\subfigure[The DAG of $mhnk, k=1$.] { 
\label{fig:mem-dag2}     
\includegraphics[width=0.65\columnwidth,trim=0 0 0 0,clip]{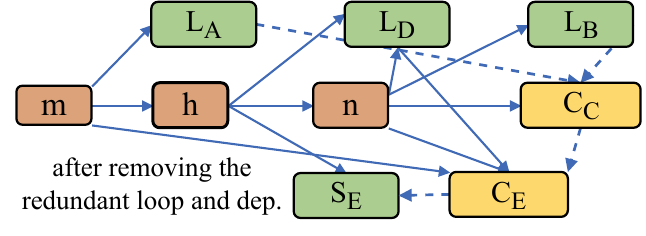}     
}  
\setlength{\abovecaptionskip}{0cm}
\caption{
DAG flow illustrating the data processing steps for the tiling expression $mhnk$ in a fused GEMM chain.
}\label{fig:mem-dag}
\vspace{-0.3cm}
\end{figure}

To illustrate this scenario, all loops and primitive statements are organized into a DAG. The dependency from loop to primitive statements is scope-dependent, indicating that the statement must be executed within the loop’s scope as the loop variables index the operands. The dependency between primitive statements is order-dependent, implying that one statement must be executed before another, but they need not be executed within the same loop’s scope. In the example of $mh(n(k(L_A, L_B, C_C), L_D, C_E), S_E)$, the DAG representation (Fig.~\ref{fig:mem-dag1}) indicates that node $L_A$ is dominated by loop node $k$, requiring it to be placed within the scope of loops $m$, $h$, $n$, $k$. Consequently, the matrix $A$ is repeatedly loaded from global memory to shared memory $hn$ times. However, when the extent of loop $k$ decreases to 1, the loop variable becomes a constant value of 0, enabling the removal of the dead node k and its corresponding dependencies from the DAG, i.e., Fig.~\ref{fig:mem-dag2}. As a result, the node $L_A$ is primarily governed by loop $m$, which enables relocating its associated memory access statement into the $m$ loop's scope (as shown in line 2 of Fig.~\ref{fig:mem-opt-code2}), thereby decreasing the access cost for $L_A$ by a factor of $hn$.

With these optimizations, MCFuser significantly reduces redundant memory accesses for each candidate in the search space, substantially easing memory bandwidth demand.

\subsection{Pruning}\label{sec:design-space-prune}

The size of the search space is determined by the number of tiling expressions multiplied by the number of candidate tile sizes. In the GEMM chain example with $M=N=1024, K=H=512$, this yields $(24+2)\times \lceil 1024/16\rceil^2 \times \lceil 512/16\rceil^2=109051904$ candidates, a substantial number. 
However, a significant portion of these candidates are either invalid or equivalent, leading to a waste of searching resources.
To efficiently explore the search space and avoid meaningless trials, we propose several guidelines to prune the search space:

\begin{figure}[tb]
\centering
\scalebox{1}{
\subfigure[The sub-tiling expression is $nk$ per thread block. ] { 
\label{fig:reduce-inner1}     
\includegraphics[width=0.45\columnwidth,trim=0 0 0 0,clip]{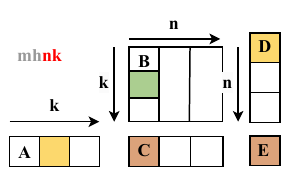}     
}  
\subfigure[The sub-tiling expression is $kn$ per thread block.] { 
\label{fig:reduce-inner2}     
\includegraphics[width=0.45\columnwidth,trim=0 0 0 0,clip]{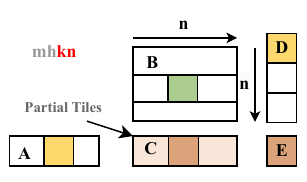}     
}  
}
\setlength{\abovecaptionskip}{0cm}
\caption{ 
Illustration of shared memory usage differences between tiling expressions $mhnk$ and $nhkn$. In (b), additional shared memory is needed to cache multiple tiles of $C$, whereas in (a), the memory footprint for a single tile can be efficiently reused.
}\label{fig:reduce-inner}
\end{figure}

\begin{itemize}
    \item {\bf Rule 1, Deduplication}: 
    We prioritize binding spatial loops with $blockIdx$ for each tiling expression, removing the loops bound to $blockIdx$ to derive the sub-tiling expression. 
    Thus, candidates sharing a sub-tiling expression per thread block are deemed equivalent. For instance, both $mhnk$ and $mnkh$ yield the same sub-tiling expression $nk$ for each thread block.
    \item {\bf Rule 2, Prevent Overwhelming Intermediate Tensor’s Shared Memory}: When the reduced loop is positioned outside the spatial loops, multiple tiles of partial results are cached in shared memory (as shown in Fig.~\ref{fig:reduce-inner}), potentially overwhelming it.
    \item {\bf Rule 3, Avoid Extra Padding}: Padding becomes necessary when the tile size does not evenly divide the dimension size. To avoid excessive padding, candidates requiring padding are discarded if the dimension size is a power of 2. For other cases, the padding ratio is kept below 0.05.
    \item {\bf Rule 4, Shared Memory Limitation}: 
    Candidates with tiles exceeding the shared memory limitation are considered illegal. We estimate the occupied shared memory as in equation~\eqref{eq:estm-shared-mem}, where $Shm_{max}$ is the maximum supported shared memory for each block per specific hardware. We prune candidates that satisfy $Shm_{estm} >1.2 Shm_{max}$, where 1.2 accommodates estimation errors. As validated in experiments (Section~\ref{sec:exp:shared-mem}), our estimation accuracy exceeds 90\%, resulting in approximately a 40\% reduction in candidates.
\end{itemize}

\begin{figure}[tb]
\centering
\scalebox{0.33}{
\centerline{\includegraphics[trim=1 1 1 0,clip]{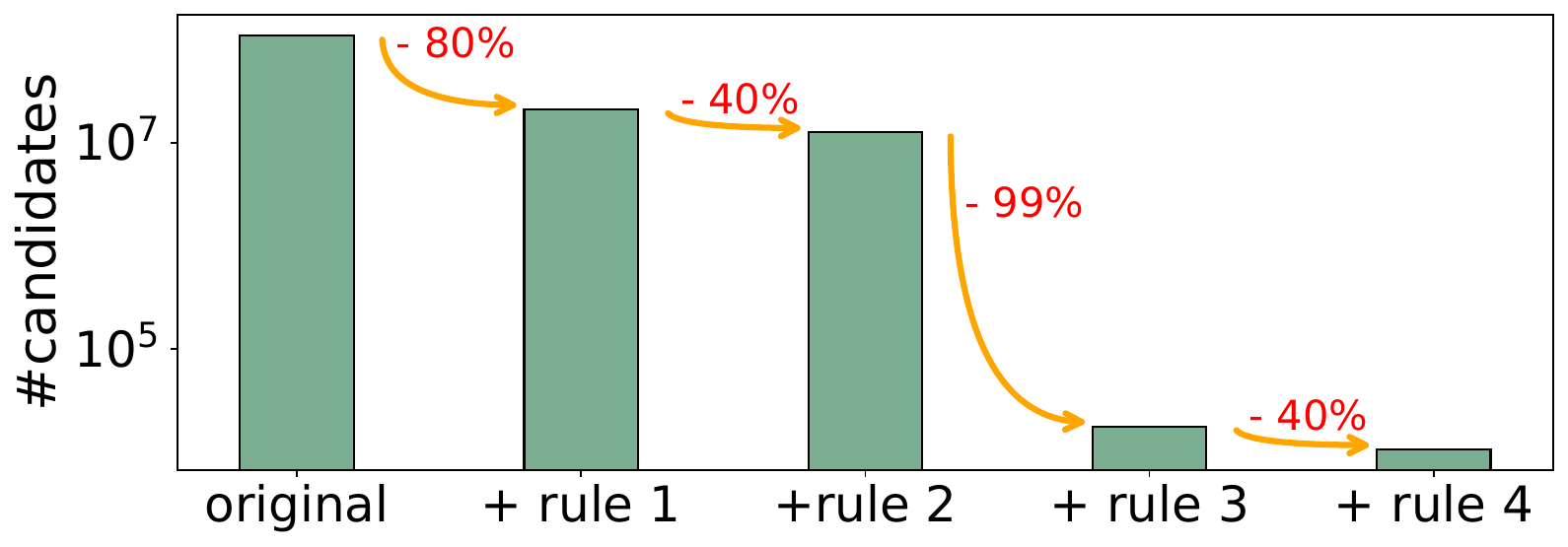}}
}
\setlength{\abovecaptionskip}{-0.1cm}
\caption{ 
The results of pruning search space candidates on an example of GEMM chain with $M=N=1024, K=H=512$.
}\label{fig:prune}
\end{figure}

\begin{equation} \label{eq:estm-shared-mem}
    Shm_{estm}=\sum_{\{Xi\}, Xi \in \mathbb{R}^{L_i\times L_j}}(T_{L_i}\times T_{L_j})
\end{equation}

Applying these rules to the GEMM chain example, we observe a drastic reduction in the search space size as illustrated in Fig.~\ref{fig:prune}. Rule 1 reduces the number of tiling expressions from 26 to 5, and Rule 2 further reduces it to 3. Rule 3 discards 99\% of impractical tile sizes, while Rule 4 eliminates 40\% of infeasible tile sizes. Ultimately, the size of the search space is reduced from $10^8$ to $10^4$ after removing trivial candidates, facilitating the exploration of optimal candidates efficiently.

\section{Performance Model and Exploration}
\label{sec:exploration}
Previously, we crafted a comprehensive and efficient search space. We now turn our attention to its exploration, starting with an analytical performance model based on a detailed analysis of computation and memory access patterns. Further, we develop a heuristic search algorithm to automatically and efficiently pinpoint the optimal scheduling.

\subsection{Analytical Performance Model} \label{sec:cost-model}
Auto tuners frequently grapple with the issue of protracted tuning times~\cite{ansor,autotvm}. To circumvent the repetitive and time-intensive performance testing, practitioners often resort to machine learning-based cost models to estimate program performance. Nonetheless, the extensive training and feature engineering required by these models can impose significant burdens on users. As a remedy, we introduce an analytical performance model that estimates the performance of generated tensor programs, detailed in equation~\eqref{eq:estm-t}. This model integrates three critical components: memory access overhead, computation overhead, and a slowdown factor, offering a streamlined approach to performance estimation.

Assuming a full schedule candidate consists of a set of computation blocks $\{C_{Xi}\}$, a set of memory access statements $\{L_{Xi}\} + \{S_{Xi}\}$, and a set of loops $(l_1,l_2,...,l_J)$, where each compute or memory access statement $St$ is surrounded by specific loops denoted as $Lp\_set(St)=(l_{p1(St)}, l_{p2(St)}, ..., l_{pk(St)})$, with $pi(St)$ representing the $i$-th surrounding loop for statement $St$. The memory access overhead~(equation~\eqref{eq:estm-mem}) for statement $St_{Xi}$ is determined by the product of the tile size $TS_{Xi}$ and the trip count $\prod_{l_j\in LP\_set(St)}l_j$, divided by the memory bandwidth $\mathcal{W}$. Similarly, the computation overhead is estimated by the product of the flops of a tile $Fp_{Xi}$ and the trip count, divided by the computation throughput $\mathcal{P}$, as in equation~\eqref{eq:estm-comp}. Additionally, the number of thread blocks affects the parallelism of fused kernels. Since each thread block is dispatched on a SM, having more thread blocks helps exploit the parallel processing power provided by the GPU. Hence, we introduce a slowdown factor in equation~\eqref{eq:estm-alpha}, where $\mathcal{N}_{SM}$ denotes the number of SMs on the GPU and $\mathcal{N}_{block}$ denotes the number of thread blocks for the fused kernel. The slowdown factor $\alpha$ decreases with the value of $\mathcal{N}_{block}$ increasing until $\alpha$ approaches 1.

Incorporating an analytical performance model into MCFuser allows us to substantially diminish the necessity for runtime measurements, thereby significantly expediting the tuning process.
Besides these, unlike approaches relying on cost models for performance estimation~\cite{ansor,flextensor}, our method eliminates the requirement for model training, thereby reducing the tuning overhead for fused kernels.
\begin{align}
    &\quad \quad \quad \quad \quad  t_{estm} = (t_{mem}+t_{comp})\times \alpha
\label{eq:estm-t} \\
    &t_{mem} =  
    \sum_{S_{Xi}\in \{L_{Xi}\}+\{S_{Xi}\}} \frac{TS_{Xi}\times\prod_{l_j\in LP\_set(S)}l_j}{\mathcal{W}} 
\label{eq:estm-mem} \\
    &\quad\   t_{comp} = 
    \sum_{S_{Xi} \in\{C_{Xi}\}}\frac{Fp_{Xi}\times\prod_{l_j\in LP\_set(S)}l_j}{\mathcal{P}} 
\label{eq:estm-comp} \\
    &\quad \quad \quad \quad \quad \alpha  = (\mathcal{N}_{block} + \mathcal{N}_{SM})/\mathcal{N}_{block}  
\label{eq:estm-alpha} 
\end{align}

\begin{algorithm}[tb]
\resizebox{0.95\linewidth}{!}{
\begin{minipage}{\linewidth}
\caption{Heuristic Search Algorithm} \label{al:search}
  \SetKwData{Up}{up} 
  \SetKwFunction{GenerateCandidates}{generateCandidates}
  \SetKwFunction{EstimatePerformance}{estimatePerformance}
  \SetKwFunction{Sort}{sort}
  \SetKwFunction{Topk}{topk}
  \SetKwFunction{Measure}{measure}
  \SetKwFunction{Topone}{top1}
  \SetKwFunction{Difference}{difference}  
  \SetKwFunction{MutatePopulation}{mutate}  
  \SetKwInOut{Input}{input}\SetKwInOut{Output}{output} 
  \SetKwInOut{Parameters}{parameters} 
    \SetAlgoLined
    \LinesNumbered 
    \Parameters{ N, n, $\epsilon$} 
    \Input{P,  the input tensor program}
    \Output{$cand\_{best}$, the best searched candidate}

    population = \GenerateCandidates(P, N) \;
    best\_t = 1e9  \;  
    best\_cand = Null \; 
    \While{True}{
        estimated\_ts = \EstimatePerformance(population) \;
        population = \Sort(population, key=estimated\_ts) \; 
        topk\_cands = \Topk(population, k=$n$, reverse=True) \; 

        topk\_ts = \Measure(topk\_cands) \; 
        top1\_t, top1\_cand = \Topone(topk\_ts, topk\_cands) \; 
        
        \If{\Difference(top1\_t, best\_t) $<$ $\epsilon$}{
            best\_cand = top1\_cand; 
            break\; 
        }
        \If{top1\_t $<$ best\_t}{
            best\_cand = top1\_cand \; 
            best\_t = top1\_t \; 
        }
        \tcc{generate next population}
        population = \MutatePopulation(population, weight=estimated\_ts) \; 
    } 
    {\bf Return} best\_cand \;
\end{minipage}
}
\end{algorithm}

\subsection{Exploration}\label{sec:explore}
Despite pruning the search space, as discussed in Section~\ref{sec:design-space-prune}, a substantial number of candidates often remain. To streamline the selection process, we've refined the evolutionary search algorithm initially developed in Ansor with two key enhancements to boost search efficiency. First, we've replaced Ansor's cost model with our analytical performance model, detailed in Section~\ref{sec:cost-model}, eliminating the time-consuming model training phase and thereby reducing computational overhead. Second, unlike Ansor, which requires manually setting the number of tuning iterations, a process that demands considerable trial and error, we have implemented a convergence criterion that automatically terminates the search, further simplifying the optimization process.

Algorithm~\ref{al:search} delineates our methodology. Initially, we randomly select $N$ candidates from the search space to establish the initial population (line 1). Utilizing our performance model from Section~\ref{sec:cost-model}, we then estimate the execution time for each candidate within this population (line 5), proceeding to isolate the top $n$ candidates (significantly fewer than $N$) for detailed measurement. 
Should the gap between the $top1\_t$ and the foremost measured result within this group fall below $\epsilon$, signifying convergence, the search process halts (lines 10-12). If not, we refresh the best result (lines 13-16) and craft the subsequent population (line 17), deriving it through mutation from its predecessor. Assuming the preceding population is ${Cand_i}$, with estimated execution times denoted by $1/et_i$ for ${i=1,2,..., N}$, each candidate ${Cand_i}$ is assigned a weight of $1/et_i$. We then randomly draw $N$ candidates from ${Cand_i}$, weighted by these values. For each selected candidate, one loop is chosen to mutate the tile size, generating a new candidate for the next population.

Through search space pruning and the integration of a heuristic search algorithm with an analytical performance model, MCFuser effectively automates and optimizes the search for the optimal fused kernel implementation.

\section{Implementation}\label{sec:impl}

MCFuser is built on top of TVM~\cite{tvm} and Triton~\cite{triton}.

\subsection{Intra-block optimization and Code Generation}
Triton~\cite{triton}, an open-source language and compiler, excels in expressing and compiling tiled neural network computations into efficient machine code. With its user-friendly interfaces, Triton enables developers to adeptly design high-performance compute kernels at the tile level, managing intra-tile optimizations autonomously. 

In MCFuser, we concentrate on leveraging Triton for inter-tile optimization, utilizing its suite of intra-tile optimizations. These include automatic coalescing, thread swizzling, automatic vectorization, tensor core-aware instruction selection, and efficient shared memory allocation and synchronization. This approach simplifies the optimization process and ensures the generated code is highly efficient.

Moreover, we harness Triton's capabilities to produce optimized PTX (Parallel Thread Execution) code\cite{ptx}, which is then seamlessly integrated with TVM, as detailed in the following subsection. This integration leverages the strengths of both Triton and TVM, creating an effective and adaptable optimization pipeline for tensor programs.

\subsection{Framework and Front-end}
To streamline our efforts and avoid redundant work, we leverage the existing compiler stack provided by TVM to the fullest extent possible. This includes utilizing the front-end languages Relay IR and TIR to express models, perform graph-level transformations, and handle runtime operators.

When processing a MBCI operator defined in Relay IR, we initially translate it into both a tiling expression and a TIR representation. These two representations are mutually convertible. The transition from a tiling expression to a TIR module is facilitated using {\tt tvm.tir.Schedule} primitives such as $tile$, $split$, $reorder$, and $bind$.  To further streamline this process, we have crafted a TIR AST visitor that extracts the tiling expression directly from a TIR module. An IR translator has also been developed to morph a TIR module into a TritonIR module. 
Finally, the PTX code produced by Triton is converted into the TVM runtime library using the {\tt runtime.module.loadfile\_ptx} interface and encapsulated within an {\tt OperatorModule} for evaluation.

When presented with a deep learning model originating from frameworks like PyTorch~\cite{pytorch}, TensorFlow~\cite{tensorflow}, or ONNX~\cite{onnx}, we initially convert it into the Relay module. Subsequently, we employ a partitioner to segment the model into MBCI sub-graphs and other components. The MBCI parts are then optimized using MCFuser, and the optimized PTX code is incorporated into the TVM runtime module. For other operators, we either continue optimization with Ansor or Relay, a decision determined flexibly based on experiments outlined in Section~\ref{sec:exp:e2e}. Finally, all runtime libraries for the operators are integrated into the executable {\tt GraphExecutorFactoryModule}.

By leveraging the capabilities of TVM's compiler stack, we ensure efficient handling of both MBCI operators and other components of deep learning models, facilitating seamless integration and optimization throughout the compilation process.

\begin{table}[tb]
\centering
\caption{The Configuration of Batch GEMM chains.}\label{tab:gemm-chain-configs}
\scalebox{0.88}{
\begin{tabular}{llllll}
\toprule
\bf Name & \bf batch & \bf M & \bf N & \bf K & \bf H \\
\toprule
G1/G2/G3 & 1 &  512 & 256 & 64 & 64/128/256 \\
G4/G5/G6 & 1 &  512 & 512 & 256/512/1024 & 256 \\
G7/G8/G9 & 1 &  512/1024/2048 & 512 & 128 &128  \\
G10/G11/G12 & 1/4/8 &  1024 & 1024 & 128 &128  \\
\bottomrule
\end{tabular}}
\label{tab:exp:gemm-chains}
\end{table}

\begin{table}[tb]
\centering
\caption{The Configurations of Self Attention Modules.}\label{tab:gemm-chain-configs}
\scalebox{0.9}{
\begin{tabular}{lllllll}
\toprule
\bf Name & \bf \#heads & \bf M & \bf N & \bf K & \bf H & Network\\
\toprule
S1 & 8 &  512 & 512 & 64 & 64 & Bert-Small\\
S2 & 12 &  512 & 512 & 64 & 64 & Bert-Base \\
S3 & 16 &  512 & 512 & 64 & 64 & Bert-Large \\
S4 & 12 &  256 & 256 & 64 & 64 & ViT-Base\\
S5 & 16 &  256 & 256 & 64 & 64 & ViT-Large\\
S6 & 16 &  256 & 256 & 80 & 80 & ViT-Huge\\
S7/S8/S9 & 1 &  512/768/1024 & 256/384/512 & 64 & 64 & MLP-Mixer\\
\bottomrule
\end{tabular}}
\label{tab:exp:selfatten}
\end{table}

\section{Evaluation}\label{sec:exp}

\begin{figure*}
\centering
\scalebox{0.36}{
\centerline{\includegraphics[trim=1 1 1 0,clip]{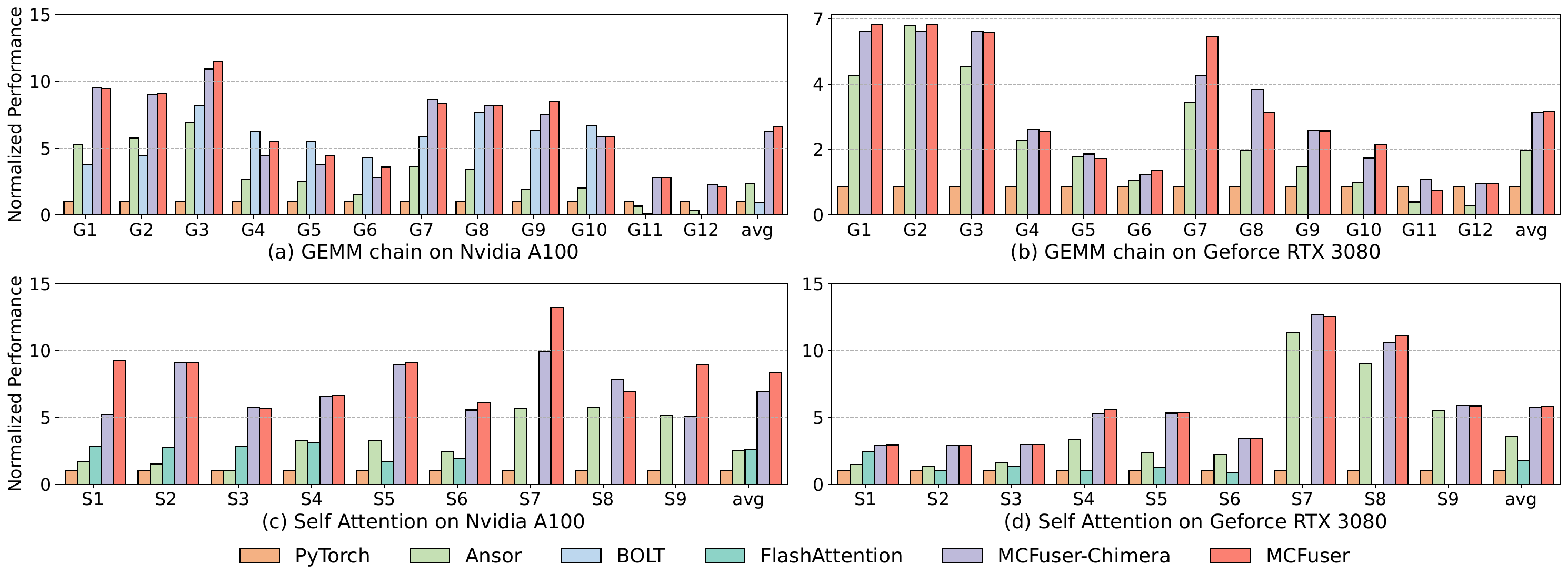}}
}
\setlength{\abovecaptionskip}{0cm}
\caption{ 
The relative performance of fusing batch GEMM chains and self-attention modules on GPUs. 
}\label{fig:exp:graph}
\end{figure*}

\subsection{Experimental Setup}

\noindent {\bf Platforms.} We evaluate MCFuser on two servers. The first is equipped with one NVIDIA A100-PCIE with 40GB and 72-core Intel(R) Xeon(R) Gold 6240C CPUs clocked at 2.60GHz. The second server features one NVIDIA GeForce RTX 3080 GPU and a 20-core Intel(R) Core(TM) i9-10900X CPU running at 3.70GHz. For simplicity, we refer to these servers as A100 and RTX3080 in the experiments. The major compilers and libraries utilized by MCFuser and other baselines include TVM 0.14, Triton 2.1, CUDA 11.7, FlashAttention\footnote{https://github.com/HazyResearch/flash-attention.git@57ee618}, Transformers 4.16\footnote{https://github.com/huggingface/transformers}, and PyTorch 2.1.

\noindent {\bf Workloads.} We assess both sub-graph fusion performance and end-to-end network performance. The sub-graphs we analyze comprise various GEMM operator chains, as well as self-attention modules from Bert~\cite{bert}, Vit~\cite{vit}, and MLP-Mixer~\cite{mlpmixer}. Additionally, we evaluate the performance of end-to-end networks on Bert models.

\noindent {\bf Comparisons.} Our baselines encompass both hand-tuned libraries and cutting-edge compilers. For libraries, we compare against PyTorch~\cite{pytorch} (utilizing CuBlas~\cite{cublas} and CuDNN~\cite{cudnn}) and FlashAttention~\cite{flashattention} (optimized for self-attention modules). Regarding compilers, we compare against leading machine learning compilers, including Relay~\cite{relay}, Ansor~\cite{ansor}, BOLT~\cite{bolt}~(built upon TVM~\cite{tvm} and cutlass~\cite{cutlass}). For Ansor, we conduct 1000 tuning trials for each subgraph. To ensure a rigorous assessment of our search space generation effectiveness against the closed-source Chimera~\cite{chimera}, we implement MCFuser-Chimera. This adaptation integrates Chimera’s search space into our framework, allowing for a direct and controlled comparison.

\subsection{Subgraph Performance}
\label{sec:exp:graph}

\subsubsection{Batch GEMM Chain} 
The configurations for the batch GEMM chains are detailed in Table~\ref{tab:exp:gemm-chains}. Here, the dimension $(batch, M, K) \times (batch, K, N)$ specifies the size for the first batch GEMM operator, while $(batch, M, N) \times (batch, N, H)$ delineates the size of the subsequent batch GEMM operator.

Figures~\ref{fig:exp:graph}(a) and (b) depict the performance of subgraphs normalized to PyTorch. For the A100 and RTX 3080 GPUs, the average speedup is 6.6$\times$ and 3.7$\times$ compared to PyTorch, 2.7$\times$ and 1.6$\times$ compared to Ansor, and 1.06$\times$ and 1.07$\times$ compared to MCFuser-Chimera. However, except for the cases where MCFuser and MCFuser-Chimera produce identical fused kernels, e.g. G1, G8 in Fig.~\ref{fig:exp:graph}(a), MCFuser achieves 1.17$\times$ and 1.13$\times$ speedups compared to MCFuser-Chimera on A100 and RTX 3080 GPUs, respectively, underscoring the efficiency of our search space generation approach.
Additionally, MCFuser delivers a 7.1$\times$ speedup over BOLT on A100. Note that BOLT does not support GPUs with \textit{sm86} compute capability, including RTX 3080. 

The speedup compared to PyTorch is attributed to fusing the memory-bound batch GEMMs and reducing off-chip memory access. Ansor adopts declarative loop-oriented scheduling but suffers from inefficient schedule space and lengthy tuning time~\cite{hidet}. Additionally, we achieve up to 5.9$\times$ and speedup compared to Ansor in the cases where Ansor fails to fuse the GEMM chains (e.g., G12 in Fig.~\ref{fig:exp:graph}(a)); this is because it concentrates more on the optimization of general operators, leading to sub-optimal performance of fusing MBCI operators. 
BOLT leverages Cutlass, the state-of-the-art open-source DNN template library for GPUs, to generate fused kernels. However, its flexibility is limited as the templates are manually developed by experts. Moreover, BOLT performs worse in some extreme cases of large tensor size, such as G11 and G12. In comparison with MCFuser-Chimera, MCFuser encompasses larger search spaces, as introduced in Section~\ref{sec:tiling-expression}, leading to better performance in various cases (e.g., G3, G4, G5 as depicted in Fig.~\ref{fig:exp:graph}(a).

\subsubsection{Self-Attention Module}
The input configurations for the self-attention modules, detailed in Table~\ref{tab:exp:selfatten}, employ the same dimensional parameters ($M$, $N$, $K$, and $H$) as used in the GEMM Chains. Unlike GEMM chains, the self-attention module includes a broader range of operators, such as softmax, which is further divisible into a series of smaller operators.

Figures~\ref{fig:exp:graph}(c) and \ref{fig:exp:graph}(d) showcase the performance of self-attention modules normalized to PyTorch. 
For the A100 and RTX 3080 GPUs, the average speedup is 8.1$\times$ and 5.8$\times$ compared to PyTorch, 2.8$\times$ and 1.45$\times$ compared to Ansor, 3.0$\times$ and 3.3$\times$ compared to FlashAttention, and 1.1$\times$ and 1.01$\times$ compared to MCFuser-Chimera, respectively. While FlashAttention employs a clever design for fusing self-attention modules, its implementation suffers from inflexibility. Specifically, a rigid constraint of $K = H$ prevents it from fusing specific modules where $K$ and $H$ differ. Additionally, FlashAttention only considers splitting the $M$ and $N$ dimensions into tiles, neglecting $K$ and $H$, which likely results in sub-optimal performance for many cases. BOLT lacks the ability to fuse self-attention modules due to its constrained fusion patterns. 

The results underscore MCFuser's capability in achieving high-performance fusion for memory-bound compute-intensive operators, validating its efficiency and effectiveness.

\subsection{End-to-End Performance}
\label{sec:exp:e2e}
For the end-to-end network performance evaluation, we utilize \textit{Bert-Small}, \textit{Bert-Base}, and \textit{Bert-Large} as workloads, setting the sequence length to 512. The key configurations of these models are listed in Table~\ref{tab:exp:selfatten}. We employ BOLTs as the baseline. Despite BOLT’s limitations in handling self-attention modules, it incorporates optimizations such as epilogue fusions (e.g., GEMM+bias+ReLU). Additionally, we compare \textit{MCFuser} with Relay and Ansor. Since \textit{MCFuser} primarily focuses on fusing memory-bound compute-intensive operators, we integrate it with Relay and Ansor to leverage their broader operator coverage. The variants of \textit{MCFuser} for end-to-end networks are denoted as \textit{MCFuser+Relay} and \textit{MCFuser+Ansor}, respectively.

\begin{figure}
\centering
\scalebox{0.32}{
\centerline{\includegraphics[trim=1 1 1 0,clip]{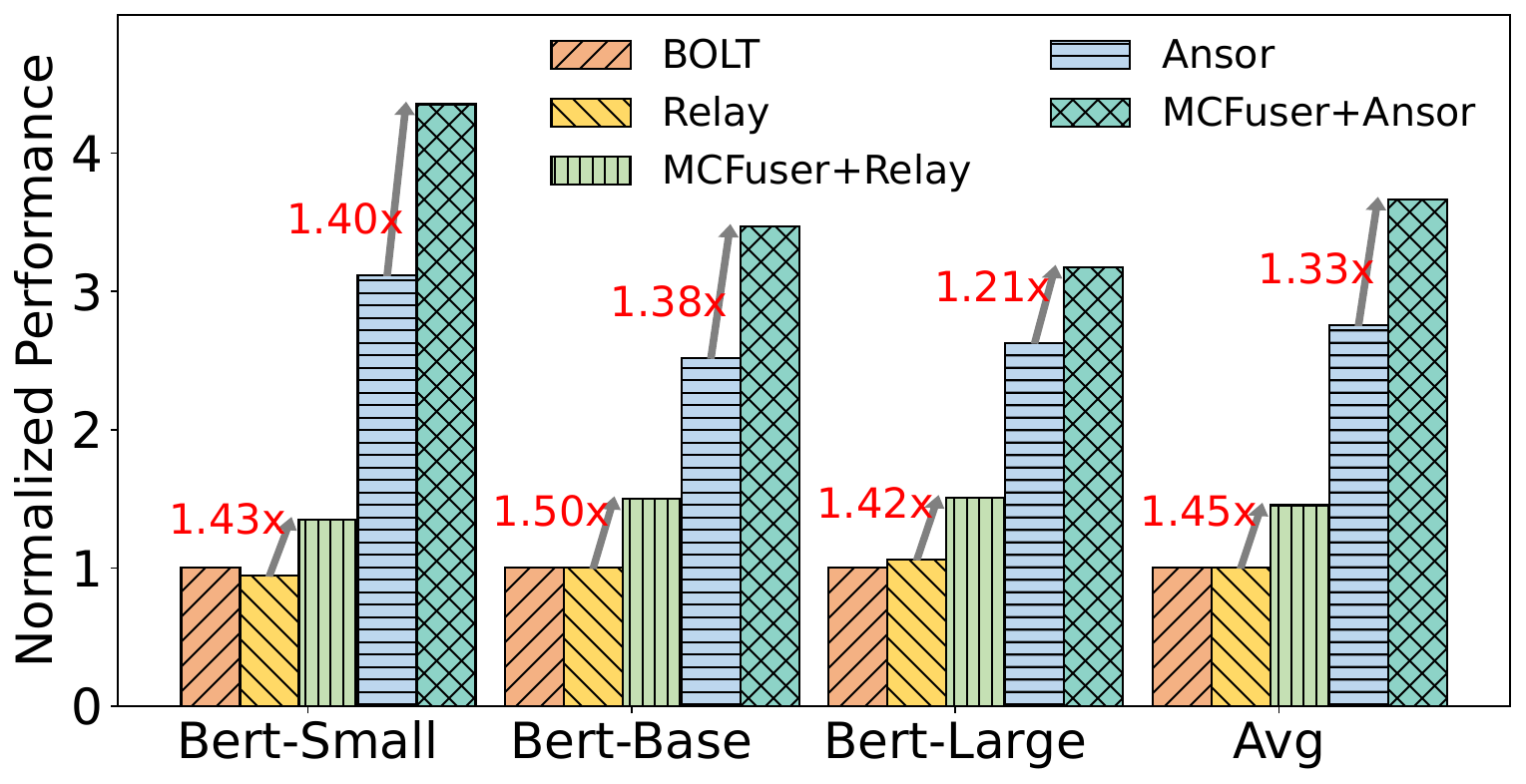}}
}
\setlength{\abovecaptionskip}{0cm}
\caption{ 
The end-to-end model evaluation on A100 GPU. 
}\label{fig:exp:e2e}
\end{figure}

Our comprehensive evaluation, performed on an A100 GPU, is depicted in Fig.~\ref{fig:exp:e2e}. It reveals that Relay, while a robust framework, delivers suboptimal performance due to its dependence on pre-defined templates without subsequent fine-tuning. Conversely, BOLT faces limitations from its pattern table, which restricts it to fusing a narrow set of operators. This approach largely leaves BOLT dependent on Relay’s implementations, resulting in only slight improvements.

In contrast, \textit{MCFuser+Relay} significantly outperforms Relay alone, achieving an impressive average speedup of 1.45×. This performance is further elevated when integrated with \textit{Ansor}, with \textit{MCFuser+Ansor} vastly outperforming \textit{BOLT} by securing an average speedup of 3.66$\times$. Remarkably, even against Ansor, which is celebrated for its highly optimized kernels, \textit{MCFuser+Relay} demonstrates superior performance, boasting an average speedup of 1.33$\times$. 
This analysis underscores the effectiveness of \textit{MCFuser} in elevating the computational efficiency across varying platforms.

\begin{table}
\centering
\caption{Tuning times for sub-graph modules and end-to-end models. "MCFuser+Relay" and "MCFuser+Ansor" columns indicate the respective speedups relative to \textit{BOLT} and \textit{Ansor}.}
\label{tab:gemm-chain-configs}
\scalebox{0.9}{
\begin{tabular}{l|c|c|c|c|c}
\toprule
\toprule
\multicolumn{6}{c}{\bf Sub Graph} \\
\toprule
 & BOLT & Ansor & \makecell{MCFuser-\\Chimera} & MCFuser & \makecell{Speedup to \\ BOLT/Ansor}\\
\midrule
GEMM Chain & 88s & 4895s & 29s & 35s & \bf 2.5$\times$~/~139$\times$\\
\midrule
Self Attention & - & 2897s & 32s & 39s &\bf -~/~74$\times$\\
\toprule
\toprule
\multicolumn{6}{c}{\bf End-to-End} \\
\toprule
methods & Relay & BOLT & \makecell{MCFuser+\\Relay} & Ansor &\makecell{MCFuser+\\Ansor}  \\
\midrule
Bert-Small& 30s  & 94s & 81s~(\textbf{1.16$\times$}) & 4.03h & 2.78h~(\textbf{1.45$\times$})\\
\midrule
Bert-Base & 149s & 227s & 202s~(\textbf{1.12$\times$}) & 3.94h & 2.74h~(\textbf{1.43$\times$})\\
\midrule
Bert-Large & 186s & 383s & 243s~(\textbf{1.57$\times$}) & 4.06h & 2.98h~(\textbf{1.36$\times$})\\

\bottomrule
\end{tabular}}
\label{tab:exp:tuning}
\end{table}

\subsection{Tuning time}
Table~\ref{tab:exp:tuning} provides detailed tuning time for both sub-graphs and end-to-end networks on the A100 GPU. When tuning sub-graph modules, MCFuser significantly outperforms BOLT (2.5$\times$ faster on average) and Ansor (139$\times$ faster for GEMM chains and 74$\times$ faster for self-attention modules on average). This speedup is attributed to MCFuser’s efficient search space and analytical performance model, which eliminates the time-consuming hardware profiling and model training required by Ansor, as well as the extensive template instantiation and massive program measurements in BOLT.

In the context of end-to-end network tuning, our first variant, \textit{MCFuser+Relay}, significantly outperforms both Relay and BOLT, as depicted in Fig.~\ref{fig:exp:e2e}. This enhancement is realized with a minimal time cost increase over Relay (under one minute) and achieves a tuning time reduction ranging from 1.12$\times$ to 1.57$\times$ compared to BOLT. Moreover, \textit{MCFuser+Ansor} attains a 1.41$\times$ average speedup in tuning time over Ansor alone. Notably, the bulk of the tuning time for \textit{MCFuser+Ansor} is consumed by operators processed by Ansor. To further reduce the tuning time, our future work will aim to expand MCFuser’s framework to include a broader array of operators.

\begin{figure}
\centering
\scalebox{0.6}{
\centerline{\includegraphics[trim=1 1 1 0,clip]{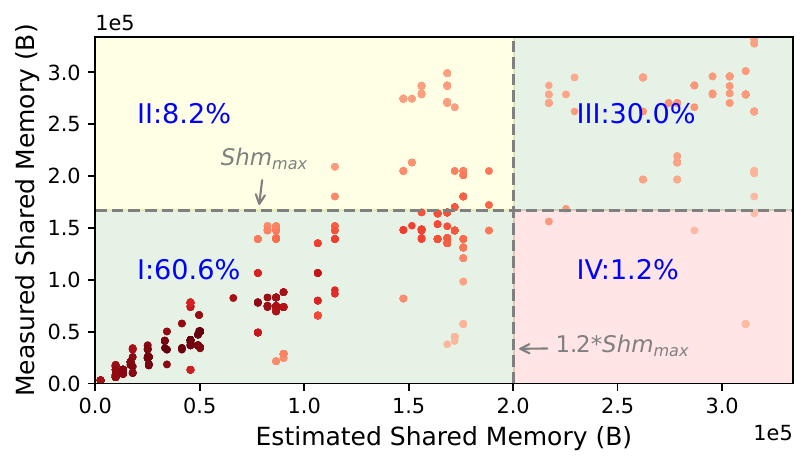}}
}
\setlength{\abovecaptionskip}{0cm}
\caption{
Model-predicted vs. actual shared memory usage on the NVIDIA A100 GPU, with $Shm_{max}$ denoting the maximum usable shared memory per thread block, factoring in special optimizations. The color shading of data points indicates data density.
}\label{fig:exp:shared_mem}
\end{figure}

\subsection{Effectiveness of the System Design}
Our system design employs Equation~\eqref{eq:estm-shared-mem} to estimate shared memory usage, effectively pruning infeasible scheduling candidates. Furthermore, an analytical performance model is utilized to forecast the performance of candidates, facilitating their selection. To validate the efficacy of these approaches, we undertake the following experimental procedures.

\subsubsection{Shared Memory Estimation}
\label{sec:exp:shared-mem}
Figure~\ref{fig:exp:shared_mem} displays the relationship between estimated shared memory usage per thread block and actual measurements performed by the NVPTX backend in LLVM, as applied to scheduled candidates from various experiments detailed in Section~\ref{sec:exp:graph}. This figure is segmented into four quadrants, distinguished by the y-axis at $Shm_{max}$, the maximal shared memory allotment per thread block, beyond which candidates are considered unexecutable on GPUs. The x-axis bifurcation at 1.2$\times Shm_{max}$, an empirically determined threshold, aims to adequately account for estimation inaccuracies. 

Notably, over 90\% of data points reside within quadrants I and III, affirming the precision of our estimation methodology in aptly identifying and pruning unviable schedules, thus curtailing the search space by 40\%. While 8.2\% of candidates (quadrant II) that should have been pruned remain, their influence on the final program performance is minimal, as they are subsequently eliminated during PTX code lowering. A mere 1.2\% of candidates are not correctly dropped despite nearing the $Shm_{max}$ limit, yet upon closer examination, these cases are inconsequential due to their considerably inferior performance compared to the optimal selection.

\begin{figure}
\centering
\scalebox{0.6}{
\centerline{\includegraphics[trim=1 1 1 0,clip]{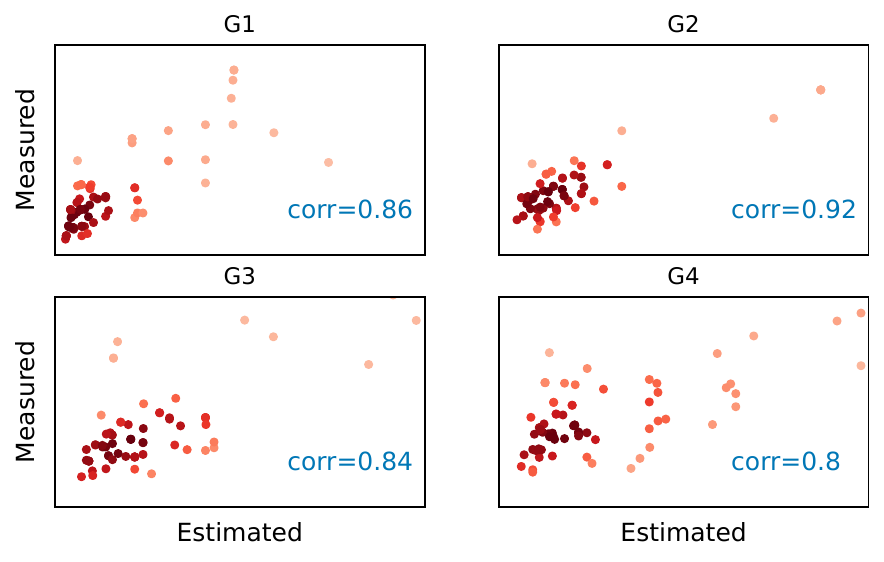}}
}
\setlength{\abovecaptionskip}{0cm}
\caption{ 
The model predicted performance versus actual measurement on NVIDIA A100 GPU. The closer to the zero point, the better the performance. The shade of the points’ color indicates
the density of the data.
}\label{fig:exp:cost_model}
\end{figure}

\subsubsection{Analytical Performance Model}
To assess our analytical performance model's accuracy, we examined the correlation between the estimated and actual performance of scheduled candidates for the first four cases listed in Table~\ref{tab:exp:gemm-chains}. As shown in Fig.\ref{fig:exp:cost_model}, the correlation coefficients  0.86, 0.92, 0.84, and 0.8 underscore the reliability of our analytical performance model. As detailed in Algorithm\ref{al:search}, the model identifies the top $n$ candidates (where $n$ is empirically set to 8) for further evaluation, enhancing our chances of identifying the optimal solution. Furthermore, the concentration of data points near the origin in Fig.~\ref{fig:exp:cost_model} highlights the effectiveness and expressiveness of our search space generation approach.

\section{Related Work} \label{sec:related}
Various hand-tuned libraries~\cite{onednn, oneapi, cublas,cudnn} and code generation compilers~\cite{tiramisu,tvm,autotvm,taso,rammer,dnnfusion,relay,astra,tensor-comprehension,scalable-kernel-fusion,akg,ansor} have been developed to enhance the performance of Deep Learning models. Among these efforts, operator fusion stands out as a prevalent optimization technique aimed at reducing kernel launch overhead and memory accesses. Depending on the type of operator and optimization method, we classify operator fusion into three categories: hand-tuned or library-based, memory-intensive, and compute-intensive.

\noindent {\bf Hand-tuned or library-based} 
Wang et al.~\cite{cross-layer-data-reuse} investigate fusing convolutional layers in CNNs, while Liang et al.~\cite{spatial-temporal-multitask} focus on maximizing hardware utilization through spatial-temporal kernel fusion. Astra~\cite{astra} is notable for fusing GEMM operators in RNNs, albeit lacking low-level code generation and relying on hand-tuned libraries. Ashari et al.~\cite{kernel-fusion2015} implement fused kernels for specific machine learning patterns, and FlashAttention~\cite{flashattention, flashattention2} provides hand-tuned optimizations for self-attention modules. While these approaches achieve high performance, their reliance on hand-tuned kernels or manual optimization limits their flexibility and adaptability to new operators. In contrast, MCFuser’s integration within a compiler framework offers a more generalizable solution for optimizing diverse operators.

\noindent {\bf Memory-Intensive Operators}
Numerous compilers~\cite{xla,halide-with-tree,tiramisu,tvm,nimble,tensor-comprehension,disc} employ fusion optimizations for memory-intensive operators. However, these techniques often struggle to address complex dependencies, resulting in incomplete fusion outcomes. Works such as Fusion Stitching~\cite{fusionstitching}, and AStitch~\cite{astitch} attempt to expand fusion scope by utilizing shared and global memory for intermediate storage. However, by treating compute-intensive operators as fusion boundaries, they overlook valuable optimization opportunities that fusing these operators would provide. In comparison, MCFuser leverages the identification of memory-bounded compute-intensive operators to push the boundaries of operator fusion, enabling more creative and efficient fusion designs.

\noindent {\bf Compute-Intensive Operators}
Optimization of compute-intensive operators has been explored by several previous works~\cite{halide-with-tree,tiramisu,tvm,generate-quantized,tensor-contraction,nimble,tensor-comprehension,ansor}. However, most of these works are limited to scenarios that fuse one compute-intensive operator with non-compute-intensive ones. Frameworks like NeoFlow~\cite{neoflow} avoid fusing multiple compute-intensive operators entirely due to code generation constraints. DNNFusion~\cite{dnnfusion}, designed for mobile devices (e.g., ARM CPU and GPU), fails to fuse compute-intensive operators because its fusion algorithm always considers fusing compute-intensive operators as non-beneficial. BOLT~\cite{bolt} relies on the Cutlass~\cite{cutlass} template library to generate code for fused GEMM, and operator types are strictly constrained by predefined pattern tables (e.g., BOLT can’t recognize einsum operators). Chimera~\cite{chimera} focuses on minimizing data movement in block execution, neglecting the impact of redundant computation, but its search space is limited by the block execution order.

In contrast, we present an exhaustive fusion strategy that generates a comprehensive search space, complemented by an analytical model that efficiently captures both memory and computation overheads for superior design choices.

\section{Conclusion}~\label{sec:conclude}
In this work, we have unveiled fusion opportunities for a novel class of operators known as MBCI operators. By developing a tiling expression that precisely defines the search space, we have facilitated the efficient fusion of these operators, thereby enhancing the generation of high-performance tensor programs. Through DAG analysis and memory access optimization, coupled with strategic pruning guidelines and the integration of an analytical performance model with a heuristic search approach, we streamlined the tuning process. Validated on NVIDIA A100 and RTX 3080 GPUs, our methods demonstrated the capability of MCFuser to significantly elevate performance, achieving up to a 5.9$\times$ speedup over Ansor, and surpassing other methods. Importantly, MCFuser also dramatically reduces tuning times by more than 70-fold, highlighting its potential to profoundly impact machine learning machine learning compilation through intelligent optimization and integration strategies.

\section*{Acknowledgment}
This work was supported by the National Key
Research and Development Program of China
(2023YFE0205700), National Natural Science Foundation
of China (62302348, 62341410), Fundamental Research
Funds for the Central Universities (2042023kf0132),
General Program of Hubei Provincial Natural Science
Foundation of China (2023AFB831), Special Fund of
Hubei Luojia Laboratory (220100016) and the Science
and Technology Development Fund (FDCT) Macau SAR
(File no. 0078/2023/AMJ). The numerical calculations in this paper have been done on the supercomputing system in the Supercomputing Center of Wuhan University.


\bibliography{main}
\bibliographystyle{bibtex/IEEEtran.bst}
\vspace{12pt}

\end{document}